# Intrinsic Frequency Limit of Direct Modulation of Resonant-Tunneling-Diode Terahertz Emitters and Effect of External Feedback Injection

Masahiro Asada[1, a)] and Safumi Suzuki [1]

[1] Department of Electrical and Electronic Engineering, Tokyo Institute of Technology,
2-12-1-S9-3 Ookayama, Meguro-ku, Tokyo 152-8552, Japan

[a)] Corresponding author: asada@pe.titech.ac.jp

**ABSTRACT**

Output power of resonant-tunneling-diode (RTD) terahertz (THz) emitters can be modulated by the bias modulation similar to a semiconductor laser. This property is useful for applications of the THz waves to wireless communications and radars. In this paper, we theoretically analyze the modulation-frequency dependence of the output response using an equivalent circuit of the RTD-THz oscillator. It is shown that there exists an intrinsic cutoff frequency of modulation in RTD, which is independent of the external circuit that supplies the modulation signal to the oscillator. This cutoff frequency is determined by the time constant given by the capacitance of RTD divided by the absolute value of negative differential conductance minus loss conductance of the oscillator, and is about 100 GHz for typical RTD-THz oscillators. We also analyze the effect of external feedback injection on the modulation characteristics. If the period of the modulation frequency is equal to an integral multiple of the round-trip time of the feedback, a dip or peak occurs in the modulation response of the output, depending on whether the THz carrier components in the output and the feedback are in phase or out of phase. We also discuss the possibility of increase in cutoff frequency by the feedback with short round-trip time.

## I. INTRODUCTION

The terahertz (THz) band with a frequency of about 0.1 to several THz is expected to have various applications, such as imaging, analysis in chemistry and biotechnology, and communication [1,2]. A compact solid-state THz source is important for these applications. For semiconductor single devices, quantum cascade lasers (QCLs) are studied [3–5] on the optical device side. Recently, room-temperature THz sources with difference frequency generation using mid-infrared QCLs has been reported [6–8]. On the electron device side, Gunn and transit-time diodes [9–11] and transistors such as heterojunction bipolar transistors (HBT), high electron mobility transistor (HEMT), and CMOS [12–16] are studied as THz sources. Recently, development of high-frequency transistors has made significant progress. Other than semiconductors, THz emitters using the intrinsic Josephson junctions in the layered high-temperature superconductor are also studied [17,18].

Resonant tunneling diode (RTD) is also one candidate for room-temperature THz sources [19–21]. Room-temperature oscillation up to 1.98 THz and relatively high output power with array configuration have been reported [22,23], and structures for higher frequency and high output power are being studied [24,25].

In the THz sources using RTD oscillators, the THz output is easily modulated by superimposing a signal on the bias voltage [26] similar to the modulation of bias current of a semiconductor laser. This direct modulation has been applied to THz wireless communications and radars [27–32]. In the experiment of the direct modulation, the upper limit of the modulation frequency is determined by that of the external circuit which supplies the modulation signal to the oscillator [26]. The intrinsic upper limit of the modulation frequency of the RTD-THz oscillator itself is not experimentally and theoretically clear. In this paper, it is theoretically shown that the RTD-THz oscillator itself has an intrinsic upper limit of modulation frequency. The effect of external feedback injection on the frequency dependence of the modulation is also analyzed.

## II. DIRECT MODULATION OF RTD

Figure 1 shows a typical structure of the RTD oscillator [21]. DC bias and modulation signal are also shown in Fig. 1. RTD is placed at the edge of one side of a slot antenna which works as a resonator and a radiator. The upper electrode of the RTD is connected to the other side of the slot through the bridge and the capacitance formed by an MIM (metal–insulator–metal) structure. This MIM structure is used to isolate the bias lines to the upper and lower electrodes of the RTD. Outside of the slot antenna and RTD, a resistor for stabilization is connected in parallel with the RTD to suppress parasitic oscillations formed by the circuit connected to the DC power supply and the modulation signal. The output power is extracted from the substrate (bottom) side through a silicon (Si) lens. Although different structures other than Fig. 1 have been reported [21,24,25], these oscillators also have the main parts that works the same as the parts in Fig. 1 described above.

RTD has a negative differential conductance (NDC) region in the current-voltage ($I_{bias}$-$V_{bias}$) curve, as shown in Fig. 2 (a). The oscillation occurs in this region, if the absolute value of NDC compensates for the loss of the slot antenna in Fig. 1. Figure 2 (b) shows the output power $P_{out}$ of the oscillator as a function of bias voltage $V_{bias}$. $P_{out}$ is generated in a range of $V_{bias}$ slightly narrower than the NDC region, which satisfies the above oscillation condition, and has the maximum value at the center of NDC [21,33].

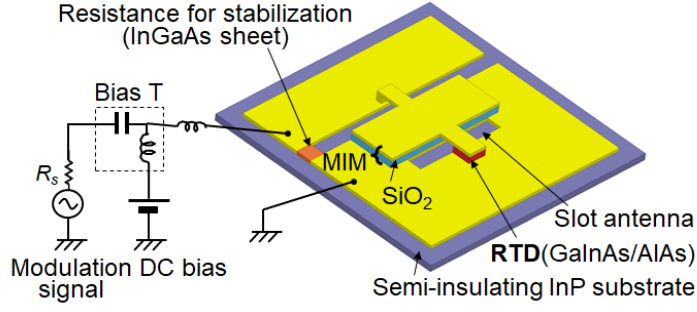

**FIG. 1.** Structure of RTD oscillator with DC bias and modulation signal.

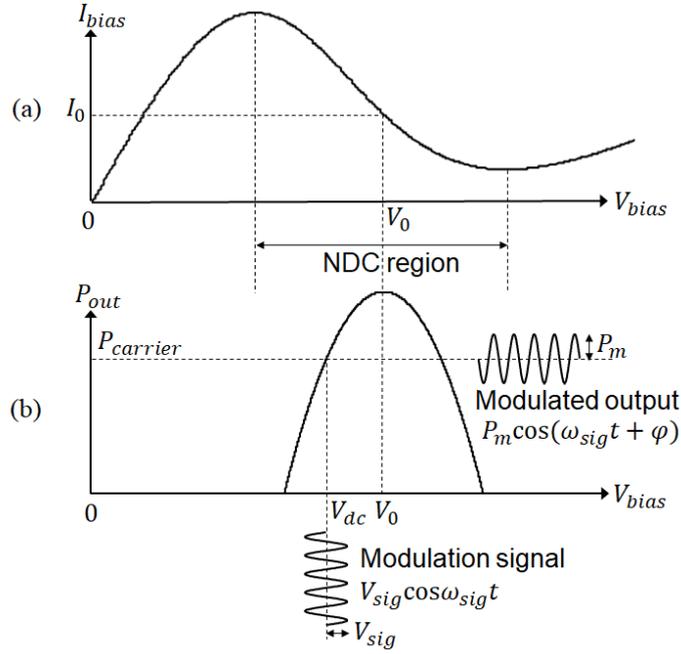

**FIG. 2.** Schematic illustration of (a) current-voltage and (b) output power-voltage characteristics of RTD.

The output power is modulated by superimposing the signal on the bias voltage, as shown in Fig. 1. As schematically shown in Fig. 2 (b), by the modulation signal $V_{sig} \cos \omega_{sig} t$ superimposed on the bias voltage $V_{bias}$, The output power is modulated as $P_m \cos(\omega_{sig} t + \varphi)$ around the carrier component $P_{carrier}$.

In the previous experiments including the modulation of RTD oscillators [26–28, 31, 32], the upper limit of the modulation frequency was determined by the circuit composed of the MIM structure and resistance for stabilization, which are outside the intrinsic part of the oscillator, and the frequency response of the intrinsic part is not clear. In this paper, the modulation response $P_m/V_{sig}$, i.e., the ratio of the amplitude $P_m$ of the modulated output power to the amplitude $V_{sig}$ of the modulation signal, is theoretically analyzed for the intrinsic part of the oscillator as a function of modulation frequency $f_{sig}$ $(= \omega_{sig}/2\pi)$.

## III. ANALYSIS OF MODULATION CHARACTERISTICS WITHOUT FEEDBACK INJECTION

First, the modulation response in the case without feedback injection is analyzed, based on the equivalent circuit of the RTD oscillator shown in Fig. 3 [21]. The external circuit outside the intrinsic part of the oscillator is not included in Fig. 3. $v_{ac}$ is the oscillation voltage across the RTD, $L$ and $C$ are the inductance and capacitance of the resonance circuit composed of the antenna and RTD, respectively, and $G_{rad}$ and $G_{loss}$ are the radiation conductance and nonradiative loss conductance of the antenna, respectively. The total conductance of the antenna is given by $G_{ant} = G_{rad} + G_{loss}$. Parasitic elements around RTD are neglected as they do not affect the conclusion.

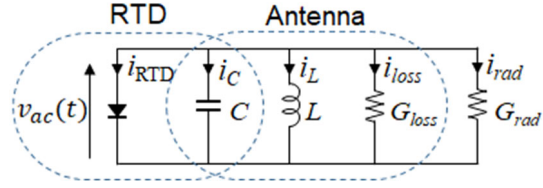

**FIG. 3.** Equivalent circuit of RTD oscillator

The relation between $I_{bias}$ and $V_{bias}$ is approximated by [33,34]

$$I_{bias}(V_{bias}) = -a(V_{bias} - V_0) + b(V_{bias} - V_0)^3 + I_0, \tag{1}$$

where $a$ and $b$ are constants, and $V_0$ and $I_0$ $(= -aV_0 + bV_0^3)$ are the voltage and current at the center of NDC. The oscillation current through RTD is calculated as

$$i_{RTD} = I_{bias}(V_{bias} + v_{ac}) - I_{bias}(V_{bias}) = -[a - 3b(V_{bias} - V_0)^2]v_{ac} + 3b(V_{bias} - V_0)v_{ac}^2. \tag{2}$$

Effect of the electron delay in RTD on NDC [35] is neglected in Eq. (2).

Substituting the current through each circuit element ($i_{RTD}$, $i_C = C\, dv_{ac}/dt$, $i_L = L^{-1}\int v_{ac}dt$, $i_{rad} = G_{rad}v_{ac}$, and $i_{loss} = G_{loss}v_{ac}$) into the equation $i_{RTD} + i_C + i_L + i_{rad} + i_{loss} = 0$ and differentiating this equation with respect to $t$, the following basic equation is obtained.

$$\frac{d^2v_{ac}}{dt^2} - \frac{\alpha - \beta v_{ac} - \gamma v_{ac}^2}{C}\frac{dv_{ac}}{dt} + \frac{1}{LC}v_{ac} = 0, \tag{3}$$

where

$$\left. \begin{array}{l} \alpha(V_{bias}) = a - G_{ant} - 3b(V_{bias} - V_0)^2, \\ \beta(V_{bias}) = 6b(V_{bias} - V_0), \\ \gamma = 3b. \end{array} \right\} \tag{4}$$

$L$, $C$, and $G_{ant}$ are assumed to be independent of applied voltage.

Under the direct modulation, the bias voltage is expressed as

$$V_{bias} = V_{dc} + V_{sig} \cos \omega_{sig} t . \tag{5}$$

$V_{dc}$ and $V_{sig} \cos \omega_{sig} t$ are schematically shown in Fig. 2 (b). $v_{ac}$ is obtained from Eqs. (3)-(5). We assume the following form for $v_{ac}$.

$$v_{ac} = V_{ac}(t) \cos \omega_{osc} t , \tag{6}$$

where $V_{ac}$ and $\omega_{osc}$ are the amplitude and angular frequency of the THz wave. If the bias voltage is not modulated, $V_{ac}$ is constant for time. Hereafter, $V_{ac}(t)$ is assumed to change much slower than $\cos \omega_{osc} t$.

Substituting Eq. (6) into Eq. (3) and neglecting the harmonic components of $\omega_{osc}$, the following equations are obtained (see Supplementary material S-I-A for the derivation).

$$\omega_{osc} = \frac{1}{\sqrt{LC}} , \tag{7}$$

$$\frac{dV_{ac}}{dt} = \frac{1}{2C} \left\{ \alpha(V_{dc}) - \frac{\gamma}{4} V_{ac}^2 - \beta(V_{dc}) V_{sig} \cos \omega_{sig} t - \gamma V_{sig}^2 \cos^2 \omega_{sig} t \right\} V_{ac} . \tag{8}$$

Assuming small signal modulation ($|V_{dc} - V_0| \gg V_{sig}$), $V_{ac}(t)$ in Eq. (8) is approximated by

$$V_{ac}(t) = V_{carrier} + V_{mod}(t) , \tag{9}$$

where $V_{carrier}$ and $V_{mod}(t)$ are the carrier and modulation components in $V_{ac}(t)$, respectively. Substituting Eq. (9) into Eq. (8), the following equations are obtained from the 0th and 1st order terms of $V_{mod}(t)$.

$$\alpha(V_{dc}) - \frac{\gamma}{4} V_{carrier}^2 = 0 , \tag{10}$$

$$\frac{dV_{mod}}{dt} + \frac{\gamma V_{carrier}^2}{4C} V_{mod} = -\frac{\beta(V_{dc}) V_{carrier}}{2C} V_{sig} \cos \omega_{sig} t . \tag{11}$$

Equation (10) gives $V_{carrier} = 2\sqrt{\alpha(V_{dc})/\gamma}$, which is equal to the oscillation voltage without modulation, and Eq. (11) gives the steady solution of $V_{mod}(t)$ in the form of $V_{mod}(t) = V_m \cos(\omega_{sig} t + \varphi)$, where $V_m$ and $\varphi$ are the voltage amplitude and phase of the modulation component given by

$$V_m = \frac{|\beta(V_{dc})| V_{carrier}}{2\alpha} \frac{1}{|1 + j\omega_{sig}/\omega_c|} V_{sig} , \tag{12}$$

$$\varphi = -\tan^{-1}\left(\frac{\omega_{sig}}{\omega_c}\right) \quad (+\pi \text{ if } V_{dc} - V_0 > 0). \tag{13}$$

$\omega_c$ in these equations stands for the cutoff angular frequency of modulation given by

$$\omega_c = \frac{\alpha(V_{dc})}{C} = \frac{a - G_{ant} - 3b(V_{dc} - V_0)^2}{C} . \tag{14}$$

The output power radiated from the slot antenna is equal to the power consumed at $G_{rad}$ in Fig. 3. The output power averaged over one period of oscillation frequency is calculated as

$$P_{out} = \frac{\omega_{osc}}{2\pi}\int_0^{\frac{2\pi}{\omega_{osc}}} G_{rad}\, v_{ac}^2(t)\, dt \simeq \frac{1}{2} G_{rad}\, V_{ac}^2(t) \simeq \frac{1}{2} G_{rad}\, V_{carrier}^2 + G_{rad} V_{carrier} V_{mod}(t). \quad (15)$$

The first and second terms on the right-hand side of Eq. (15) are the carrier and modulation components ($P_{carrier}$ and $P_{mod}(t) = P_m \cos(\omega_{sig} t + \varphi)$ in Fig. 2 (b)), respectively, which are expressed as

$$P_{carrier} = 2G_{rad}\frac{\alpha(V_{dc})}{\gamma} = 2G_{rad}\left\{\frac{a - G_{ant}}{3b} - (V_{dc} - V_0)^2\right\}, \quad (16)$$

$$P_{mod}(t) = P_m \cos(\omega_{sig} t + \varphi) = \frac{4G_{rad}|V_0 - V_{dc}|}{|1 + j\,\omega_{sig}/\omega_c|} V_{sig} \cos(\omega_{sig} t + \varphi). \quad (17)$$

From Eq. (17), the modulation response $P_m/V_{sig}$ is obtained as

$$\frac{P_m}{V_{sig}} = \frac{4G_{rad}|V_0 - V_{dc}|}{|1 + j\,\omega_{sig}/\omega_c|}. \quad (18)$$

At $\omega_{sig} = \omega_c$, the modulation response decreases to $1/\sqrt{2}$ of that at $\omega_{sig} \to 0$. As seen from Eq. (14), the cutoff angular frequency $\omega_c$ is equal to the reciprocal of the time constant which is given by the capacitance divided by the absolute value of NDC at the bias point $V_{dc}$ minus the loss of antenna. $P_m/V_{sig}$ at $\omega_{sig} \to 0$ is equal to $\partial P_{carrier}/\partial V_{dc}$, which is the slope of the output power with respect to bias voltage in Fig. 2 (b). The phase $\varphi$ of the modulation component is inverted if $V_{dc} - V_0 > 0$, as in Eq. (13), which corresponds to setting the bias point at the right-hand side of the peak of the output power in Fig. 2 (b).

The modulation component of the output power vanishes at $V_{dc} = V_0$ (the center of the NDC region in Fig. 2(a)), as seen from Eq. (18), in the present small signal analysis. This is because $\partial P_{carrier}/\partial V_{dc} = 0$ at this bias point (= the peak of the output power), as shown in Fig. 2 (b). At this bias point, the small signal analysis in which the last term in the right-hand-side parentheses of Eq. (8) is neglected cannot be applied, and the modulation component consists only of the harmonics higher than the second order.

In the above analysis, the modulation component $V_{mod}(t)$ of the oscillation voltage was first calculated, and then the modulation component $P_{mod}(t)$ of the output power was calculated using $V_{mod}(t)$. It is also possible to directly obtain $P_{mod}(t)$ by finding a differential equation for the output power $P_{out}$ first. Since this method complicates the mathematical procedure in the analysis for the case with feedback injection described later, the above method is used throughout this paper.

The cutoff frequency of modulation is roughly estimated as follows, assuming the same parameter values as those in Ref. 33; $a - G_{loss} = $ 10 mS, $b = $ 37 mS / V², and $G_{rad} = $ 2 mS. Although the loss conductance $G_{loss}$ is neglected in Ref. 33, it can be included by replacing $a$ in Ref. 33 with $a - G_{loss}$. At the bias point $V_0 - V_{dc} = \Delta V/4 = $ 0.125 V, $\alpha(V_{dc})$ is calculated as $\alpha(V_{dc}) = a - G_{ant} - 3b(V_{dc} - V_0)^2 = $ 6.3 mS from Eq. (4). The cutoff frequency of modulation is estimated to be $\omega_c/2\pi = $ 100 GHz for $C = $ 10 fF [34] from Eq. (14). The cutoff frequency is large for small $G_{loss}$ and small $C$.

In the reported experiments [26–28, 31, 32], the upper limit of the modulation frequency is limited by the cutoff frequency of the external circuit that supplies the modulation signal. The intrinsic cutoff frequency of RTD obtained above is higher than the cutoff frequency of the external circuit, and has not yet been observed experimentally. The intrinsic cutoff frequency of modulation will appear by improvements in the external circuit for the modulation signal.

## IV. ANALYSIS OF MODULATION CHARACTERISTICS WITH FEEDBACK INJECTION

For the analysis of the modulation characteristics with feedback injection, we consider a system shown in Fig. 4, where the output is extracted from a window with small reflection. The spherical surface of the Si lens is assumed to be anti-reflection coated except for the case it is the window. The reflection at the interface between the oscillator substrate and the Si lens is neglected as the difference of the refractive index is small between them.

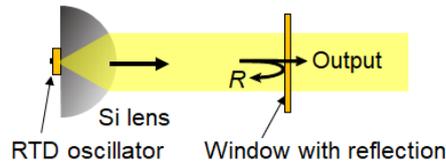

**FIG. 4.** Oscillator with feedback injection

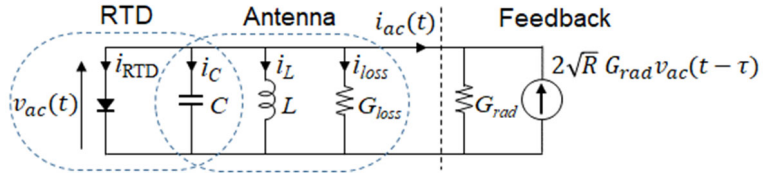

**FIG. 5.** Equivalent circuit of RTD oscillator with feedback injection (small reflection case)

The equivalent circuit of Fig. 4 is shown in Fig. 5 (see Supplementary material S-II for the derivation), where the current source equivalently expresses the feedback injection due to the reflection at the window, $R$ is the power reflectivity at the window, and $\tau$ is the round-trip time of the feedback wave between RTD and window. The equivalent circuit in Fig. 5 is derived with the assumption $R \ll 1$ (see Supplementary material S-III for the case of large $R$).

The basic equation is obtained by adding a term of the feedback injection to the right-hand side of Eq. (3) as

$$\frac{d^2 v_{ac}(t)}{dt^2} - \frac{\alpha - \beta v_{ac}(t) - \gamma v_{ac}^2(t)}{C} \frac{dv_{ac}(t)}{dt} + \frac{1}{LC} v_{ac}(t) = 2\sqrt{R}\frac{G_{rad}}{C}\frac{dv_{ac}(t-\tau)}{dt}. \quad (19)$$

Same as the case without feedback, the bias voltage is modulated as in Eq. (5) and $v_{ac}$ in Eq. (6) is also assumed here. However, since the voltage amplitude $V_{ac}(t)$ and the oscillation angular frequency $\omega_{osc}$ change due to the feedback injection, Eqs. (7) and (8) no longer hold. Again, $V_{ac}(t)$ is assumed to change much slower than $\cos\omega_{osc}t$. In addition, the change of $\omega_{osc}$ due to the feedback injection is assumed to be much smaller than $\omega_{osc}$. Substituting Eq. (6) into Eq. (19) and calculating in the same manner as in the case without feedback, the following equations are obtained instead of Eqs. (7) and (8) (see Supplementary material S-I-B for the derivation).

$$\omega_{osc} \simeq \frac{1}{\sqrt{LC}} - \frac{G_{rad}}{C}\frac{V_{ac}(t-\tau)}{V_{ac}(t)}\sqrt{R}\sin\omega_{osc}\tau, \tag{20}$$

$$\frac{dV_{ac}(t)}{dt} - \frac{1}{2C}\left\{\alpha(V_{dc}) - \frac{\gamma}{4}V_{ac}^2(t) - \beta(V_{dc})V_{sig}\cos\omega_{sig}t - \gamma V_{sig}^2\cos^2\omega_{sig}t\right\}V_{ac}(t)$$
$$\simeq \frac{G_{rad}}{C}V_{ac}(t-\tau)\sqrt{R}\cos\omega_{osc}\tau. \tag{21}$$

Equation (20) coincides with the oscillation angular frequency with feedback injection [33] in the absence of modulation ($V_{ac}(t) =$ constant in time).

Similar to the case without feedback injection, the following equations are obtained under the small-signal approximation by substituting Eq. (9) into Eq. (21).

$$\alpha(V_{dc}) - \frac{\gamma}{4}V_{carrier}^2 = -2G_{rad}\sqrt{R}\cos\omega_{osc}\tau, \tag{22}$$

$$\frac{dV_{mod}(t)}{dt} + \frac{\gamma V_{carrier}^2}{4C}V_{mod}(t) + \frac{G_{rad}}{C}\sqrt{R}\cos\omega_{osc}\tau\left\{V_{mod}(t) - V_{mod}(t-\tau)\right\}$$
$$= -\frac{\beta(V_{dc})V_{carrier}}{2C}V_{sig}\cos\omega_{sig}t. \tag{23}$$

From these equations, the carrier component $V_{carrier}$ in $V_{ac}(t)$ is obtained as

$$V_{carrier} = 2\sqrt{\frac{\alpha(V_{dc})}{\gamma}}\left(1 + 2\frac{G_{rad}}{\alpha(V_{dc})}\sqrt{R}\cos\omega_{osc}\tau\right)^{1/2}, \tag{24}$$

and the modulation component $V_{mod}(t)$ is obtained as $V_{mod}(t) = V_m\cos(\omega_{sig}t + \varphi)$ with the voltage amplitude $V_m$ and phase $\varphi$ given by

$$V_m = \frac{|\beta(V_{dc})|V_{carrier}}{2\alpha}\frac{1}{\left|1 + j\frac{\omega_{sig}}{\omega_c} + \frac{G_{rad}}{\omega_c C}\sqrt{R}\cos\omega_{osc}\tau\left(1 - e^{-j\omega_{sig}\tau}\right)\right|}V_{sig}, \tag{25}$$

$$\varphi = -\mathrm{Arg}\left\{1 + j\frac{\omega_{sig}}{\omega_c} + \frac{G_{rad}}{\omega_c C}\sqrt{R}\cos\omega_{osc}\tau\left(1 - e^{-j\omega_{sig}\tau}\right)\right\} \quad (+\pi \text{ if } V_{dc} - V_0 > 0). \tag{26}$$

Although $\omega_c$ in Eqs. (25) and (26) is given in Eq. (14) again, the frequency dependence of the modulation component is not determined only by $\omega_c$.

The output power radiated from the window in Fig. 4 is equal to the net power flowing from the left-hand side to the right-hand side of the vertical dotted line in the equivalent circuit of Fig. 5. Using $v_{ac}(t)$ and $i_{ac}(t)$ in Fig. 5, this power averaged over one period of the oscillation frequency is calculated as

$$P_{out} = \frac{\omega_{osc}}{2\pi} \int_0^{\frac{2\pi}{\omega_{osc}}} v_{ac}(t)\, i_{ac}(t)\, dt$$

$$\simeq \frac{1}{2} G_{rad}\left(1 - 2\sqrt{R} \cos \omega_{osc}\tau\right) V_{carrier}^2$$

$$+ G_{rad} V_{carrier} \left[ V_{mod}(t) - \sqrt{R} \cos \omega_{osc}\tau \{ V_{mod}(t) + V_{mod}(t-\tau) \} \right]. \tag{27}$$

The first term in the right-hand side of Eq. (27) is the carrier component $P_{carrier}$ of the output power, which is calculated using Eq. (24) as

$$P_{carrier} = 2\frac{G_{rad}}{\gamma}\left(1 - 2\sqrt{R} \cos \omega_{osc}\tau\right)\{\alpha(V_{dc}) + 2G_{rad}\sqrt{R} \cos \omega_{osc}\tau\ \}. \tag{28}$$

The second term in the right-hand side of Eq. (27) is the modulation component of the output power, which is written as $P_{mod} = P_m \cos(\omega_{sig} t + \varphi + \varphi')$ with

$$P_m = G_{rad} V_{carrier} V_m \left| 1 - \sqrt{R} \cos \omega_{osc}\tau\ (1 + e^{-j\omega_{sig}\tau}) \right|$$

$$= 4 G_{rad} |V_0 - V_{dc}| \left(1 + 2\frac{G_{rad}}{\omega_c C} \sqrt{R} \cos \omega_{osc}\tau\right)$$

$$\times \left| \frac{1 - \sqrt{R} \cos \omega_{osc}\tau\ (1 + e^{-j\omega_{sig}\tau})}{1 + j\frac{\omega_{sig}}{\omega_c} + \frac{G_{rad}}{\omega_c C} \sqrt{R} \cos \omega_{osc}\tau\ (1 - e^{-j\omega_{sig}\tau})} \right| V_{sig}, \tag{29}$$

$$\varphi' = \mathrm{Arg}\left\{ 1 - \sqrt{R} \cos \omega_{osc}\tau\ (1 + e^{-j\omega_{sig}\tau}) \right\}, \tag{30}$$

and $\varphi$ in Eq. (13). From Eq. (29), the modulation response under the feedback injection is given by

$$\frac{P_m}{V_{sig}} = 4 G_{rad} |V_0 - V_{dc}| \left(1 + 2\frac{G_{rad}}{\omega_c C} \sqrt{R} \cos \omega_{osc}\tau\right) \left| \frac{1 - \sqrt{R} \cos \omega_{osc}\tau\ (1 + e^{-j\omega_{sig}\tau})}{1 + j\frac{\omega_{sig}}{\omega_c} + \frac{G_{rad}}{\omega_c C} \sqrt{R} \cos \omega_{osc}\tau\ (1 - e^{-j\omega_{sig}\tau})} \right|. \tag{31}$$

The feedback injection changes the frequency dependence and magnitude of the modulation response. The magnitude of the modulation response is different from Eq. (17) even at $\omega_{sig} \to 0$, because the bias-voltage dependence of the output power changes due to the feedback injection.

Figure 6 shows the modulation response in the low frequency limit ($\omega_{sig} \to 0$) as a function of $\sqrt{R} \cos \omega_{osc}\tau$ normalized by the modulation response for $R = 0$. $R$ and the phase $\omega_{osc}\tau$ of the feedback wave are included in the modulation response in the form of $\sqrt{R} \cos \omega_{osc}\tau$ if $R \ll 1$. (In the case of large $R$, the modulation response also has the term including only $R$ independently, as shown in Supplementary material S-III.)

The result in Fig. 6 is understood from the second term in the right-hand side of Eq. (27) which is the original form of the modulation component $P_{mod}$. In Eq. (27), $\sqrt{R}\cos\omega_{osc}\tau$ is included in two parts; $V_{carrier}$ and the second term of the square brackets. The former increases $P_{mod}$ with increasing $\sqrt{R}\cos\omega_{osc}\tau$ as seen from Eq. (24), while the latter decreases $P_{mod}$. The characteristics of Fig. 6 are determined by these two factors. These two factors are originated from the reduction of radiation conductance by the feedback injection (see Supplementary material S-II and Ref. 33). Due to the decrease in radiation conductance, the oscillation voltage increases through the decrease in the antenna loss (the former), while the power consumption in the radiation conductance decreases (the latter).

In the numerical example given in the previous section, $G_{rad}/\omega_c C \simeq 0.32$. For the curve with $G_{rad}/\omega_c C$ close to this value in Fig. 6, the modulation response decreases or increases depending on whether the feedback wave is in phase ($\cos\omega_{osc}\tau > 0$) or in the opposite phase ($\cos\omega_{osc}\tau < 0$) with respect to the output wave. If $G_{rad}/\omega_c C$ ($= G_{rad}/\alpha(V_{dc})$) is large and $\sqrt{R}\cos\omega_{osc}\tau$ is negative with large absolute value, the oscillation does not occur, and there is no solution for $V_{carrier}$ in Eq. (24).

Figure 7 shows the frequency dependence of the modulation response normalized by that of $R = 0$ at $\omega_{sig} \to 0$. The sign of $\cos\omega_{osc}\tau$ is changed by slightly changing the round-trip time $\tau$ of the feedback injection around 100ps. This value of $\tau$ corresponds to a distance of 1.5 cm between the oscillator and the reflection point in air or 5 mm in a medium with refractive index of $n = 3$.

As shown in Fig. 6, the modulation response with feedback injection is located at the lower or upper side of the response without feedback injection ($\sqrt{R}\cos\omega_{osc}\tau = 0$), depending on whether $\cos\omega_{osc}\tau$ is positive or negative. This is originated from the result in Fig. 6.

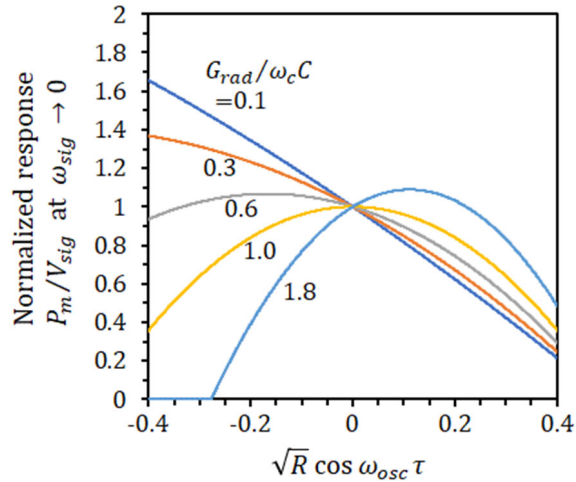

**FIG. 6.** Modulation response at low frequency limit ($\omega_{sig} \to 0$) normalized by the result at $R = 0$.

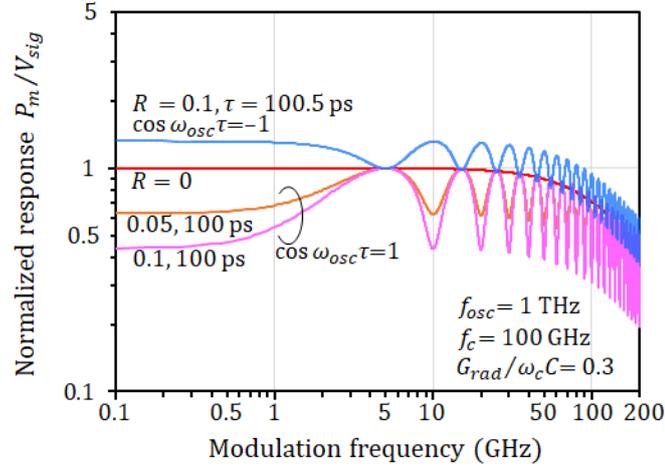

**FIG. 7.** Frequency dependence of modulation response normalized by the result at $R = 0$ and $\omega_{sig} \to 0$.

When $\cos \omega_{osc} \tau$ is positive, dips occur periodically at modulation frequencies equal to an integral multiple of $1/\tau$, and when $\cos \omega_{osc} \tau$ is negative, these dips change to peaks. These results are also understood from the second term of Eq. (27), which is the original form of the modulation component $P_{mod}$. $V_{mod}(t) + V_{mod}(t - \tau)$ in the square brackets in this part is maximized when the modulation component of the feedback wave is in phase with that of the output wave. This corresponds to the modulation frequency being an integral multiple of $1/\tau$, and it becomes a dip or peak depending on the sign of its coefficient $\cos \omega_{osc} \tau$ in Eq. (27).

The modulation response changing periodically with the phase of the feedback wave has the same origin as that for the mode locking of a laser, although the analysis in this paper using a linear approximation of small signal modulation is not applicable to pulse generation in the mode locking including a nonlinear effect.

The measured result of the modulation response similar to the above theoretical result has been reported [26]. In this measured result, a fine periodic structure is seen in the frequency dependence of the modulation response below the cutoff frequency, although the cutoff frequency of modulation is determined by the external circuit. Although various factors such as impedance matching between RTD and external circuit can be considered in the periodic structure of the measured result, the feedback injection can also be one of the causes.

Figure 8 shows the frequency dependence of the modulation response for small $\tau$, i.e., the case where the reflection point is close to the oscillator. $\tau = 2.5$ and 3 ps in Fig. 8 correspond to the distance of 125 and 150 μm between RTD and the reflection point, respectively, in a medium with refractive index of $n = 3$. This value of the refractive index is close to that of the oscillator substrate. In this example, the reflection point is assumed to be closer to the oscillator than the spherical surface of the Si lens. This condition can be realized, e.g., by placing a film, which is sufficiently thinner than the THz wavelength and has a refractive index different from that of the oscillator substrate and the Si lens, under the oscillator substrate. The period $1/\tau$ in the modulation response is 330 GHz for $\tau = 3$ ps and 400 GHz for $\tau = 2.5$ ps, which are much higher than the cutoff frequency (100 GHz) and out of range in Fig. 8.

As shown in Fig. 8, the short-range feedback makes the modulation response flat to hundreds of GHz at the expense of slightly smaller modulation response in the low frequency range.

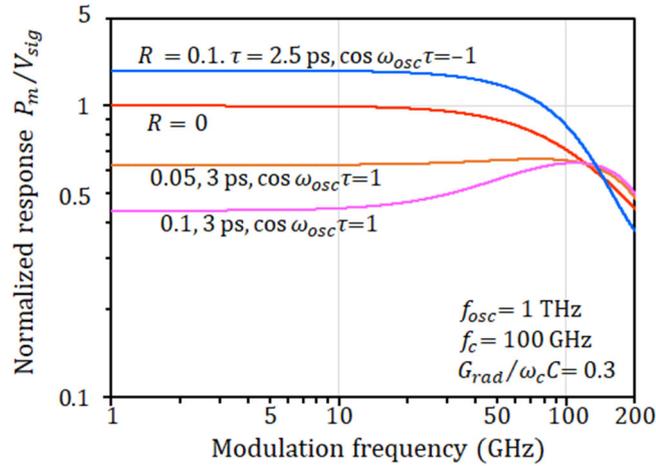

**FIG. 8.** Frequency dependence of modulation response normalized by the result at $R = 0$ and $\omega_{sig} \to 0$ for the case of short distance between the oscillator and reflection point.

## V. SUMMARY

The modulation characteristics of the RTD-THz oscillator was analyzed using an equivalent circuit. It was shown that there exists an intrinsic cutoff frequency of modulation in RTD, which is independent of the external circuit that supplies the modulation signal to the oscillator. This cutoff frequency is determined by the time constant given by the capacitance of RTD divided by the absolute value of negative differential conductance minus loss conductance of the oscillator, and is about 100 GHz for typical RTD-THz oscillators. The effect of external feedback injection on the modulation characteristics was also analyzed. If the period of the modulation frequency is equal to an integral multiple of the round-trip time of the feedback, a dip or peak occurs in the modulation response of the output, depending on whether the THz carrier components in the output and the feedback are in phase or out of phase. The feedback injection adversely affects the modulation characteristics if the reflection point is at a distance sufficiently longer than the wavelength of the modulation frequency, but may also have a possibility of flattening the frequency response of the modulation if the distance to the reflection point is comparable to or less than the wavelength of the cutoff frequency.

## ACKNOWLEDGEMENTS

The authors thank Honorary Prof. Y. Suematsu, Emeritus Profs. K. Furuya and S. Arai, Profs. Y. Miyamoto and N. Nishiyama, and Assoc. Prof. M. Watanabe of the Tokyo Institute of Technology for continuous encouragement. This work was supported by JSPS (21H04552), JST-ACCEL (JPMJMI17F2), JST-CREST (JPMJCR1534 and JPMJCR21C4), SCOPE (JP215003005), commissioned research from NICT (No. 03001), and Canon Foundation.

**SUPPLEMENTARY MATERIAL**

Contents:

S-I. Derivation of equations for the amplitude of the oscillation voltage

    S-1-A. The case without feedback injection (Eqs. (7) and (8) in the paper)

    S-1-B. The case with feedback injection (Eqs. (20) and (21) in the paper)

S-II. Derivation of equivalent circuit in Fig. 5 in the paper

S-III. Analysis without the approximation of $R \ll 1$

**S-I. Derivation of equations for the amplitude of the oscillation voltage**

**S-1-A. The case without feedback injection (Eqs. (7) and (8) in the paper)**

The first and second derivatives of Eq. (6) in the paper are approximated as follows, $V_{ac}$ changes very slowly with time compared to $\cos \omega_{osc} t$ (i.e., $\omega_{osc} \gg (dV_{ac}/dt)/V_{ac}$).

$$\frac{dv_{ac}}{dt} \simeq -\omega_{osc} V_{ac} \sin \omega_{osc} t \tag{S-1}$$

$$\frac{d^2 v_{ac}}{dt^2} \simeq -\omega_{osc}^2 V_{ac} \cos \omega_{osc} t - 2\frac{dV_{ac}}{dt} \omega_{osc} \sin \omega_{osc} t , \tag{S-2}$$

Substituting these equations into Eq. (3) in the paper, we obtain

$$\left(\frac{1}{LC} - \omega_{osc}^2\right) V_{ac} \cos \omega_{osc} t - 2\frac{dV_{ac}}{dt} \omega_{osc} \sin \omega_{osc} t$$

$$+ \frac{1}{C}\left(\alpha - \beta V_{ac} \cos \omega_{osc} t - \gamma V_{ac}^2 \cos^2 \omega_{osc} t \right) \omega_{osc} V_{ac} \sin \omega_{osc} t = 0 . \tag{S-3}$$

The factors $\cos^2 \omega_{osc} t \cdot \sin \omega_{osc} t$ and $\cos \omega_{osc} t \cdot \sin \omega_{osc} t$ included in the expansion of Eq. (S-3) are transformed to $(\sin \omega_{osc} t + \sin 3\omega_{osc} t)/4$ and $(1/2)\sin 2\omega_{osc} t$, respectively. We neglect the second and third harmonic components of $\omega_{osc}$ in the expansion terms. In the next step, Eq. (S-3) is summarized in an equation in which the sum of two terms proportional to $\cos \omega_{osc} t$ and $\sin \omega_{osc} t$ is equal to zero. Then, by equating each of these two terms to zero, Eq. (7) in the paper is obtained from the term of $\cos \omega_{osc} t$, and the following equation is obtained from the term of $\sin \omega_{osc} t$.

$$\frac{dV_{ac}}{dt} = \frac{1}{2C}\left(\alpha - \frac{\gamma}{4} V_{ac}^2\right) V_{ac} . \tag{S-4}$$

$\alpha$ in Eq. (S-4) is a function of the bias voltage as in Eq. (4) in the paper. From Eqs. (4) and (5) in the paper, we obtain

$$\alpha(V_{dc} + V_{sig} \cos \omega_{sig} t) = \alpha(V_{dc}) - 6b(V_{dc} - V_0)V_{sig} \cos \omega_{sig} t - 3b(V_{sig} \cos \omega_{sig} t)^2$$

$$= \alpha(V_{dc}) - \beta(V_{dc})V_{sig} \cos \omega_{sig} t - \gamma V_{sig}^2 \cos^2 \omega_{sig} t . \tag{S-5}$$

Equation (8) in the paper is obtained from Eqs. (S-4) and (S-5).

### S-1-B. The case with feedback injection (Eqs. (20) and (21) in the paper)

Substituting Eq. (6) into Eq. (19) in the paper, the left-hand side of Eq. (19) is calculated with the same manner as Supplementary material S-I-A. The right-hand side of Eq. (19) is transformed to

$$2\sqrt{R}\frac{G_{rad}}{C}\frac{dv_{ac}(t-\tau)}{dt} \simeq -2\sqrt{R}\frac{G_{rad}}{C}V_{ac}(t-\tau)\,\omega_{osc}\sin(\omega_{osc}t - \omega_{osc}\tau)$$

$$= 2\sqrt{R}\frac{G_{rad}}{C}V_{ac}(t-\tau)\omega_{osc}\{\cos\omega_{osc}t \sin\omega_{osc}\tau - \sin\omega_{osc}t \cos\omega_{osc}\tau\ \}. \tag{S-6}$$

Neglecting the harmonic components of $\omega_{osc}$ as in Supplementary material S-I-A, we obtain the following equations.

$$\frac{1}{LC} - \omega_{osc}^2 = 2\sqrt{R}\frac{G_{rad}}{C}\frac{V_{ac}(t-\tau)}{V_{ac}(t)}\omega_{osc} \sin\omega_{osc}\tau , \tag{S-7}$$

$$\frac{dV_{ac}(t)}{dt} - \frac{1}{2C}\left\{\alpha - \frac{\gamma}{4}V_{ac}^2(t)\right\}V_{ac}(t) = \sqrt{R}\frac{G_{rad}}{C}V_{ac}(t-\tau)\cos\omega_{osc}\tau. \tag{S-8}$$

The left-hand side of Eq. (S-7) is approximated as follows, assuming that the difference between $\omega_{osc}$ and $1/\sqrt{LC}$ is much smaller than $\omega_{osc}$.

$$\frac{1}{LC} - \omega_{osc}^2 = \left(\frac{1}{\sqrt{LC}} - \omega_{osc}\right)\left(\frac{1}{\sqrt{LC}} + \omega_{osc}\right) \simeq \left(\frac{1}{\sqrt{LC}} - \omega_{osc}\right)2\omega_{osc} \tag{S-9}$$

Equation (20) in the paper is obtained from Eqs. (S-7) and (S-9), and Eq. (21) is obtained by using Eq. (S-5) for $\alpha$ in Eq. (S-8).

### S-II. Derivation of equivalent circuit in Fig. 5 in the paper

Figure S-1 (a) shows an equivalent circuit of RTD oscillator including radiation space. First, we discuss an oscillator without modulation, which is oscillating with a single frequency, using the complex number representation. In Fig. S-1 (a), the right-hand side from the terminal pair 1-1' is represented by the transmission line which has the characteristic conductance equal to the radiation conductance $G_{rad}$ of the antenna. Although the characteristic conductance of the actual radiation space is different from $G_{rad}$, the antenna plays the role of impedance conversion. Assuming that the reflection coefficient of the electric field in the incident wave at the window is expressed as $r = \sqrt{R}\, e^{-j\phi}$, where $R$ and $\phi$ are the power reflectivity and the phase delay of the reflection, respectively, the reflection coefficient of the oscillation voltage at 1-1' is given by the transmission-line theory as $re^{-2jk_0\ell}$ with $k_0$ and $\ell$ being the wave number of the radiated electromagnetic wave and the distance between the oscillator and window (optical distance converted into air).

The admittance of the right-hand side of 1-1' is given by

$$Y_{rad} = G_{rad} \frac{1 - re^{-2jk_0\ell}}{1 + re^{-2jk_0\ell}} = G_{rad} - 2G_{rad} \frac{re^{-2jk_0\ell}}{1 + re^{-2jk_0\ell}} \tag{S-10}$$

$$\simeq G_{rad} - 2G_{rad}\, re^{-2jk_0\ell}\ (r \ll 1). \tag{S-11}$$

Equation (S-11) is an approximate formula for $r \ll 1$, which is used in the paper and Ref. 33. Using Eq. (S-11), the equivalent circuit is transformed from Fig. S-1 (a) to (c) via (b), where the angular frequency $\omega_0$ of the radiated electromagnetic wave and the round-trip time $\tau$ between the oscillator and window are used instead of $k_0$, $\ell$, and $\phi$ by the relation $2k_0\ell + \phi = \omega_0\tau$.

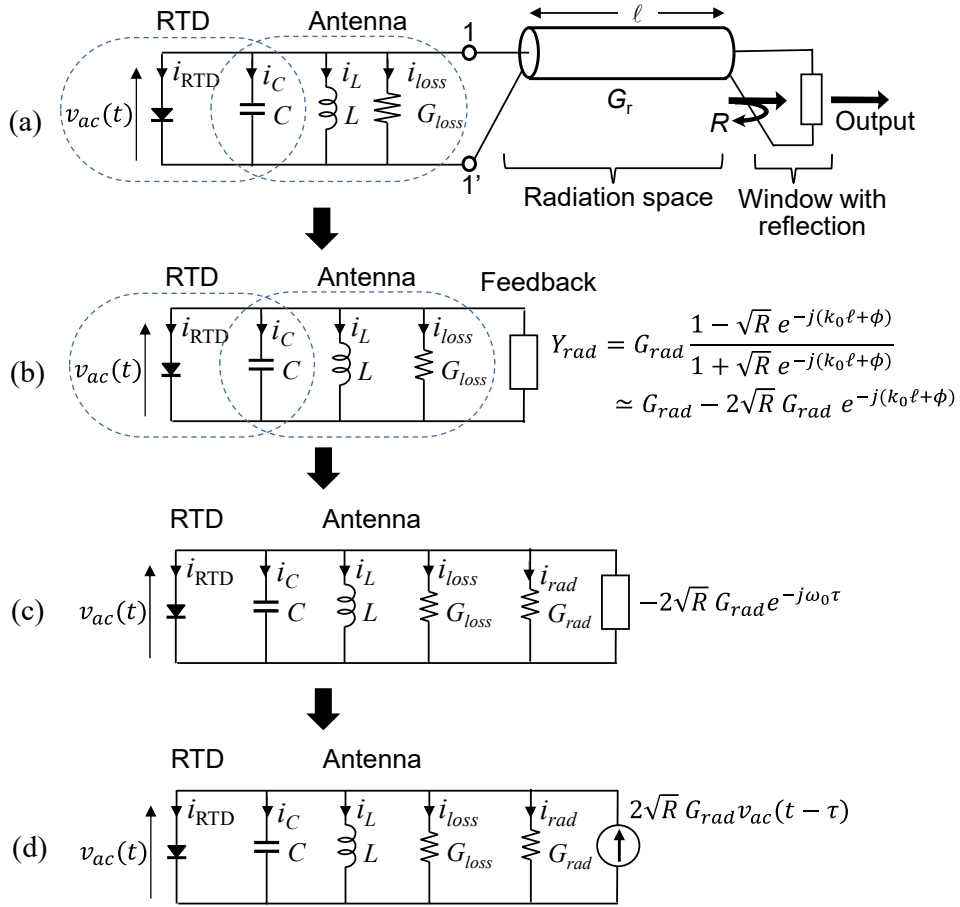

**FIG. S-1.** Equivalent circuit of RTD oscillator with feedback injection ($R \ll 1$).

Since the voltage across the admittance at the right end of Fig. S-1 (c) is equal to $v_{ac}(t)$, the current through this admittance is written as $-2\sqrt{R}G_{rad}e^{-j\omega_0\tau} \cdot v_{ac}(t)$. Since $v_{ac}(t)$ is expressed as $V_{ac}e^{j\omega_0 t}$ in the complex number representation, the current through this admittance is expressed as $-2\sqrt{R}G_{rad}e^{-j\omega_0\tau} \cdot V_{ac}e^{j\omega_0 t} = -2\sqrt{R}G_{rad}v_{ac}(t-\tau)$. The last formula holds not only for $v_{ac}(t)$ with a single frequency but also for $v_{ac}(t)$ with an arbitral temporal variation. Using this result, the admittance at the right end of Fig. S-1 (c) is replaced with a current source shown in Fig. S-1 (d), which has the upward current direction. Thus, the equivalent circuit in Fig. 4 in the paper is obtained. This current source expresses the situation in which the output wave radiated from the oscillator is reflected and reinjected into the oscillator with a delay time of $\tau$.

### S-III. Analysis without the approximation of $R \ll 1$

The analysis for large reflectivity at the window is shown here. Although the basic features of the effect of the feedback injection are well expressed in the analysis with $R \ll 1$ in the paper, precise analysis without limit for the reflectivity may be necessary for some actual cases as well as for checking the accuracy of the approximation method.

For a large reflectivity, Eq. (S-10) in Supplementary material S-II must be used instead of Eq. (S-11). Equation (S-10) is transformed as follows.

$$Y_{rad} = G_{rad} - 2G_{rad}\frac{\sqrt{R}e^{-j\omega_0\tau}}{1+\sqrt{R}e^{-j\omega_0\tau}} = G_{rad} + 2G_{rad}\sum_{n=1}^{\infty}(-\sqrt{R})^n e^{-jn\omega_0\tau} \quad (S-12)$$

The equivalent circuit is expressed as Fig. S-2 (a) instead of Fig. S-1 (c). The admittance at the right end of Fig. S-2 (a) is replaced with a current source as shown in Figure S-2 (b) by the same process as in Supplementary material S-II.

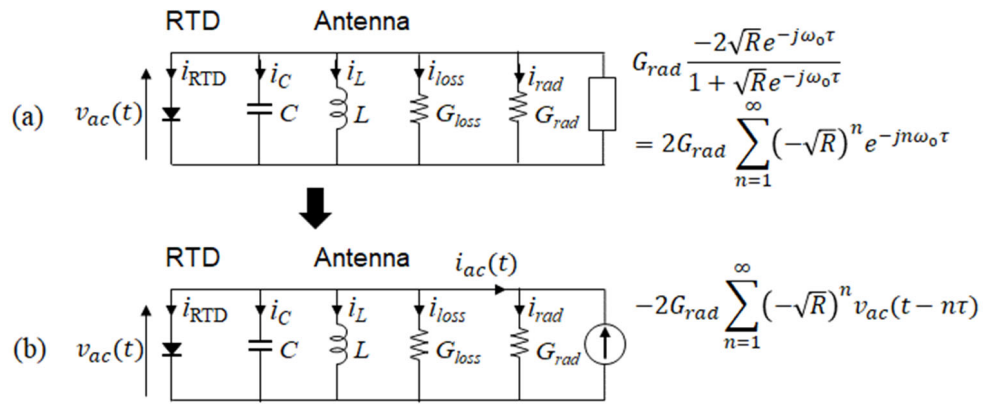

FIG. S-2. Equivalent circuit of RTD oscillator with feedback injection without approximation of $R \ll 1$.

According to Fig. S-2 (b), the basic equation is written as

$$\frac{d^2v_{ac}(t)}{dt^2} - \frac{\alpha - \beta v_{ac}(t) - \gamma v_{ac}^2(t)}{C}\frac{dv_{ac}(t)}{dt} + \frac{1}{LC}v_{ac}(t) = -2\frac{G_{rad}}{C}\sum_{n=1}^{\infty}(-\sqrt{R})^n\frac{dv_{ac}(t-n\tau)}{dt}. \quad (S-13)$$

Substituting Eq. (6) into Eq. (S-13) and calculating with the process in Supplementary material S-I, we obtain

$$\omega_{osc} \simeq \frac{1}{\sqrt{LC}} + \frac{G_{rad}}{C}\sum_{n=1}^{\infty}(-\sqrt{R})^n\frac{V_{ac}(t-n\tau)}{V_{ac}(t)}\sin n\omega_{osc}\tau, \quad (S-14)$$

$$\frac{dV_{ac}(t)}{dt} - \frac{1}{2C}\left\{\alpha(V_{dc}) - \frac{\gamma}{4}V_{ac}^2(t) - \beta(V_{dc})V_{sig}\cos\omega_{sig}t - \gamma V_{sig}^2\cos^2\omega_{sig}t\right\}V_{ac}(t)$$

$$\simeq -\frac{G_{rad}}{C}\sum_{n=1}^{\infty}(-\sqrt{R})^n V_{ac}(t-n\tau)\cos n\omega_{osc}\tau. \quad (S-15)$$

For the absence of modulation ($V_{ac}(t) = $ constant in time), Eq. (S-14) is further calculated as

$$\omega_{osc} \simeq \frac{1}{\sqrt{LC}} - \frac{G_{rad}}{C}\frac{\sqrt{R}\sin\omega_{osc}\tau}{1 + R + 2\sqrt{R}\cos\omega_{osc}\tau}. \quad (S-16)$$

Equation (S-16) gives the exact change in oscillation angular frequency due to feedback injection for the case without modulation, and it reduces to Eq. (20) in the paper if $R \ll 1$.

Substituting Eq. (9) into Eq. (S-15) and assuming small signal modulation, the following equation is obtained for the 0th order of $V_{mod}(t)$.

$$\alpha(V_{dc}) - \frac{\gamma}{4}V_{carrier}^2 = -2G_{rad}\Omega_0, \quad (S-17)$$

where

$$\Omega_0 = -\sum_{n=1}^{\infty}(-\sqrt{R})^n\cos n\omega_{osc}\tau = \sqrt{R}\frac{\sqrt{R} + \cos\omega_{osc}\tau}{1 + R + 2\sqrt{R}\cos\omega_{osc}\tau}. \quad (S-18)$$

The equation for the 1st order of $V_{mod}(t)$ is also obtained as

$$\frac{dV_{mod}(t)}{dt} + \frac{\gamma V_{carrier}^2}{4C}V_{mod}(t) - \frac{G_{rad}}{C}\sum_{n=1}^{\infty}\{V_{mod}(t) - V_{mod}(t-n\tau)\}(-\sqrt{R})^n\cos n\omega_{osc}\tau$$

$$= -\frac{\beta(V_{dc})V_{carrier}}{2C}V_{sig}\cos\omega_{sig}t. \quad (S-19)$$

The carrier component $V_{carrier}$ of the amplitude of oscillation is obtained from Eq. (S-17) as

$$V_{carrier} = 2\left\{\frac{\alpha(V_{dc}) + 2G_{rad}\Omega_0}{\gamma}\right\}^{1/2}, \quad (S-20)$$

and the voltage amplitude $V_m$ and the phase $\varphi$ in the modulation component $V_{mod}(t) = V_m\cos(\omega_{sig}t + \varphi)$ are obtained from Eq. (S-19) as

$$V_m = \frac{|\beta(V_{dc})| V_{carrier}}{2\alpha} \frac{1}{\left|1 + j\frac{\omega_{sig}}{\omega_c} + \frac{G_{rad}}{\omega_c C}(\Omega_0 - \Omega_1)\right|} V_{sig}, \qquad (S-21)$$

$$\varphi = -\text{Arg}\left\{1 + j\frac{\omega_{sig}}{\omega_c} + \frac{G_{rad}}{\omega_c C}(\Omega_0 - \Omega_1)\right\} \quad (+\pi \text{ if } V_{dc} - V_0 > 0), \qquad (S-22)$$

where $\Omega_0$ is given by Eq. (S-18), and $\Omega_1$ is given by

$$\begin{aligned}\Omega_1 &= \sum_{n=1}^{\infty}(-\sqrt{R})^n e^{-jn\omega_{sig}\tau} \cos n\omega_{osc}\tau \\ &= \frac{\sqrt{R}}{2}\left\{\frac{\sqrt{R} + e^{-j(\omega_{osc}+\omega_{sig})\tau}}{1 + R + 2\sqrt{R}\cos(\omega_{osc}+\omega_{sig})\tau} + \frac{\sqrt{R} + e^{j(\omega_{osc}-\omega_{sig})\tau}}{1 + R + 2\sqrt{R}\cos(\omega_{osc}-\omega_{sig})\tau}\right\}.\end{aligned} \qquad (S-23)$$

Similar to Eq. (27) in the paper, the output power radiated from the window in Fig. 4 is given by

$$\begin{aligned}P_{out} &= \frac{\omega_{osc}}{2\pi}\int_0^{\frac{2\pi}{\omega_{osc}}} v_{ac}(t) i_{ac}(t) dt = \frac{\omega_{osc}}{2\pi}\int_0^{\frac{2\pi}{\omega_{osc}}} v_{ac}(t)\left\{G_{rad}v_{ac}(t) + 2G_{rad}\sum_{n=1}^{\infty}(-\sqrt{R})^n v_{ac}(t - n\tau)\right\} dt \\ &\simeq \frac{1}{2}G_{rad} V_{ac}^2(t) + G_{rad}V_{ac}(t)\sum_{n=1}^{\infty}(-\sqrt{R})^n V_{ac}(t - n\tau)\cos n\omega_{osc}\tau \\ &\simeq \frac{1}{2}G_{rad}\left(1 + 2\sum_{n=1}^{\infty}(-\sqrt{R})^n \cos n\omega_{osc}\tau\right) V_{carrier}^2 \\ &\quad + G_{rad}V_{carrier}\left[V_{mod}(t) + \sum_{n=1}^{\infty}(-\sqrt{R})^n \cos n\omega_{osc}\tau \left\{V_{mod}(t) + V_{mod}(t - n\tau)\right\}\right] \\ &= P_{carrier} + P_{mod}(t). \end{aligned} \qquad (S-24)$$

The first term of the last equation in Eq. (S-24) is the carrier component given by

$$P_{carrier} = 2G_{rad}\frac{1 - R}{1 + R + 2\sqrt{R}\cos\omega_{osc}\tau}\left(\frac{\alpha(V_{dc}) + 2G_{rad}\Omega_0}{\gamma}\right). \qquad (S-25)$$

The second term is the modulation component, which is written as $P_{mod} = P_m \cos(\omega_{sig}t + \varphi + \varphi')$ with

$$\begin{aligned}P_m &= G_{rad}V_{carrier}V_m |1 - (\Omega_0 + \Omega_1)| \\ &= 4G_{rad}|V_0 - V_{dc}|\left(1 + 2\frac{G_{rad}}{\omega_c C}\Omega_0\right)\frac{|1 - (\Omega_0 + \Omega_1)|}{\left|1 + j\frac{\omega_{sig}}{\omega_c} + \frac{G_{rad}}{\omega_c C}(\Omega_0 - \Omega_1)\right|} V_{sig},\end{aligned} \qquad (S-26)$$

$$\varphi' = \text{Arg}\{1 - (\Omega_0 + \Omega_1)\}, \qquad (S-27)$$

and $\varphi$ given by Eq. (13) in the paper. From Eq. (S-26), the modulation response is given by

$$\frac{P_m}{V_{sig}} = 4G_{rad}|V_0 - V_{dc}|\left(1 + 2\frac{G_{rad}}{\omega_c C}\Omega_0\right)\frac{|1 - (\Omega_0 + \Omega_1)|}{\left|1 + j\frac{\omega_{sig}}{\omega_c} + \frac{G_{rad}}{\omega_c C}(\Omega_0 - \Omega_1)\right|}. \qquad (S-28)$$

Compared to the result for $R \ll 1$ in Eq. (31) in the paper, $\sqrt{R}\cos\omega_{osc}\tau$ and $\sqrt{R}e^{-j\omega_{sig}\tau}\cos\omega_{osc}\tau$ are replaced with $\Omega_0$ in Eq. (S-18) and $\Omega_1$ in Eq. (S-23), respectively. If $R \ll 1$, Eq. (S-28) reduces to Eq. (31).

Figures S-3 and S-4 show the calculation results of Eq. (S-28) together with the results of Eq. (31) for comparison. Figure S-3 shows the modulation response in the low frequency limit ($\omega_{sig} \to 0$), and Fig. S-4 shows the frequency dependence of the modulation response, both of which are normalized by the result for $R = 0$ at $\omega_{sig} \to 0$. It is seen from these results that the analysis assuming $R \ll 1$ gives a good approximation of the modulation response with $R$ of up to about 0.1. In Fig. S-3, the curve of the approximate analysis for $\cos\omega_{osc}\tau = 1$ reaches 0 at $R = 0.25$. At $R > 0.25$, the curve becomes negative (i.e., the phase is inverted), showing unnatural behavior in spite of the low frequency limit ($\omega_{sig} \to 0$). This obviously shows that the approximation is inaccurate in this range, and thus, the range of $R > 0.25$ on this curve is deleted in Fig. S-3.

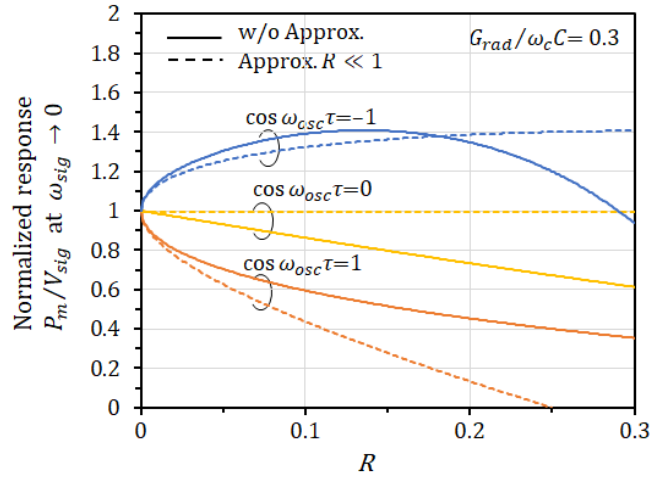

FIG. S-3. Dependence of modulation response on reflectivity at $\omega_{sig} \to 0$ normalized by the result without feedback injection.

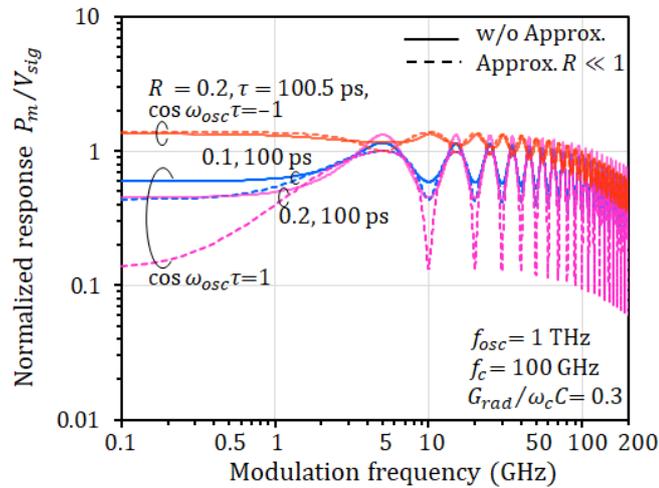

FIG. S-4. Frequency dependence of modulation response normalized by the result at $\omega_{sig} \to 0$ without feedback injection.